\documentclass[twocolumn,superscriptaddress,aps,prl]{revtex4-1}
\usepackage{amssymb,amsmath}
\usepackage{graphicx}
\usepackage[colorlinks,linkcolor=blue,citecolor=blue,urlcolor=blue]{hyperref}

\begin{document}

\title{Fractional Brownian motion with reflecting wall}
\author{Alexander H. O. Wada}
\affiliation{Department of Physics, Missouri University of Science and Technology,
Rolla, MO 65409, USA}
\affiliation{Instituto de F\'isica, Universidade de S\~ao Paulo, Rua do Mat\~ao, 1371,\\ 05508-090 S\~ao Paulo, S\~ao Paulo, Brazil}

\author{Thomas Vojta}
\affiliation{Department of Physics, Missouri University of Science and Technology,
Rolla, MO 65409, USA}
\affiliation{Kavli Institute for Theoretical Physics, University of California, Santa Barbara, CA 93106-4030, USA}

\begin{abstract}
Fractional Brownian motion, a stochastic process with long-time
correlations between its increments, is a prototypical model for anomalous diffusion.
We analyze fractional Brownian motion in the presence of a reflecting wall by means of
Monte Carlo simulations. While the mean-square displacement of the particle
shows the expected anomalous diffusion behavior $\langle x^2 \rangle \sim t^\alpha$,
the interplay between the geometric confinement and the long-time memory
leads to a highly non-Gaussian probability density function with a power-law singularity
at the barrier. In the superdiffusive case, $\alpha> 1$, the particles accumulate at the barrier
leading to a divergence of the probability density. For subdiffusion, $\alpha < 1$,
in contrast, the probability density is depleted close to the barrier.
 We discuss implications of these findings, in particular
for applications that are dominated by rare events.
\end{abstract}

\date{\today}

\maketitle

\paragraph*{Introduction.}
\label{sec:Intro}

Diffusion is a ubiquitous phenomenon with applications in physics, chemistry,
biology, and many other fields. Normal diffusion is characterized by a linear
dependence of the mean-square displacement $\langle x^2 \rangle$ of the moving particle
on the time $t$. Within the probabilistic approach pioneered by Einstein
\cite{Einstein_book56}, it can be understood as a stochastic process that is
local in time and space. This means that (i) the motion features a finite correlation time
beyond which individual displacements can be considered independent random variables,
and (ii) the displacements during a correlation time have a finite second moment.

If at least one of these conditions is not fulfilled, deviations from the linear time
dependence $\langle x^2 \rangle \sim t$ may appear, i.e., the diffusion may be anomalous.
The list of systems in which subdiffusive motion (for which $\langle x^2 \rangle$ grows
slower than $t$) or superdiffusive motion (where $\langle x^2 \rangle$ grows faster than $t$)
have been experimentally observed is extensive; and different mathematical models have
been developed to account for these measurements (for reviews see, e.g., Refs.\
\cite{MetzlerKlafter00,HoeflingFranosch13,BressloffNewby13,MJCB14,MerozSokolov15,MetzlerJeonCherstvy16,Norregaardetal17}
and references therein). Anomalous diffusion
is currently reattracting  considerable attention because modern microscopic techniques
give unprecedented access to the motion of single molecules in complex environments
\cite{XCLLL08,BrauchleLambMichaelis12,ManzoGarciaParajo15}.

A possible mechanism leading to anomalous diffusion consists in long-range
power-law correlations in time between individual displacements (steps).
The prototypical model for this situation is fractional Brownian motion
(FBM) \cite{Kolmogorov40,MandelbrotVanNess68}, a non-Markovian self-similar Gaussian process
with stationary increments. The mean-square displacement of FBM follows the power
law $\langle x^2 \rangle \sim t^\alpha$. It is characterized by the anomalous diffusion
exponent $\alpha$ \footnote{In the mathematical literature, the Hurst exponent $H=\alpha/2$
is often used instead of $\alpha$.} that can take values between 0 and 2.
In the subdiffusive case, $0<\alpha<1$, the increments are anticorrelated (antipersistent)
while the motion is persistent (positive correlations between the steps) in the
superdiffusive case $1<\alpha<2$. The marginal case, $\alpha=1$, separating the two
regimes  corresponds to normal Brownian motion with uncorrelated increments.

FBM has been studied extensively in the mathematical literature (see, e.g.,
Refs.\ \cite{Kahane85,Yaglom87,Beran94,BHOZ08}). It has found applications
in diverse fields of science and beyond, including, for example, polymer dynamics \cite{ChakravartiSebastian97,Panja10},
diffusion inside living cells \cite{SzymanskiWeiss09}, traffic in electronic networks
\cite{MRRS02}, as well as the dynamics of stock markets (see, e.g., Ref.\ \cite{RostekSchoebel13}
and references therein).
 Nonetheless, many of its properties remain
poorly understood, in particular in the presence of nontrivial boundary conditions
(an exception is the first-passage behavior on a semi-infinite domain \cite{HansenEngoyMaloy94,DingYang95,KKMCBS97,Molchan99}).
This is related to the fact that a description of FBM at the level of a (generalized)
diffusion equation has not yet been found, and the method of images to solve boundary
value problems does not apply \footnote{Note that diffusion equations with a time-dependent
diffusion constant that are sometimes written down in connection with FBM actually
describe a fundamentally different process called scaled Brownian motion \cite{LimMuniady02,JeonChechkinMetzler14},
see the concluding section of this paper for a discussion.}.

Here, we focus on a paradigmatic example of FBM in a confined
geometry, viz., one-dimensional FBM in the presence of a reflecting wall or barrier that restricts the motion
to the nonnegative $x$-axis.
We perform large-scale Monte Carlo simulations of a discrete-time
version of FBM \cite{Qian03} covering the superdiffusive and subdiffusive regimes.
We find that the mean-square displacement $\langle x^2\rangle$ of a particle that starts at
the origin shows the expected $t^\alpha$ time dependence, just as in the free, unconfined case.
However, due to the interplay of the long-range correlations and the confinement,
the probability density function $P(x,t)$ of the particle position
features surprising, highly non-Gaussian behavior.
In the superdiffusive regime, $\alpha> 1$, the particles accumulate at the barrier.
This leads a divergence of the probability density for $x\to 0$. The subdiffusive
regime, $\alpha < 1$, features the opposite behavior. The particles are depleted near the barrier,
and the probability density goes to zero for $x\to 0$. Both singularities are well
described by power laws. In the remainder of the paper, we introduce the model, describe
our simulations, and discuss in detail their results as well as implications of our findings.

\paragraph*{Discrete-time FBM.}

We employ the discrete-time FBM described by Qian \cite{Qian03}.
Consider a free (unconfined) particle that starts at the origin at time $t=0$.
Its total displacement $x_t$ at integer time $t$  is the result of a sequence
of discrete steps, $x_t= x_{t-1} + \xi_t$. The increments $\xi_i$
constitute a fractional Gaussian noise, i.e., they are identical Gaussian random variables
of zero mean, variance $\sigma^2$, and correlation function
\begin{equation}
 C(j) = \langle \xi_i \xi_{i+j} \rangle = \frac 1 2 \sigma^2 \left(|j-1|^\alpha -2|j|^\alpha +|j+1|^\alpha \right) ~.
\label{eq:xi_corr}
\end{equation}
In the long-time limit, $j\to\infty$, the correlations take the power-law form
$ \langle \xi_i \xi_{i+j} \rangle \sim  \alpha(\alpha-1) j^{\alpha-2}$. They are
positive (persistent) for $\alpha>1$ and negative (antipersistent) for $\alpha<1$.
The resulting correlation function of the displacements is easily evaluated; it reads
\begin{equation}
 \langle x_s x_t \rangle = \frac 1 2 \sigma^2 \left(s^\alpha -|s-t|^\alpha +t^\alpha \right) ~.
\label{eq:x_corr}
\end{equation}
For $s=t$, this implies anomalous diffusion with mean-square displacement
$\langle x_t^2 \rangle = \sigma^2 t^\alpha$.

To implement the reflecting wall at $x=0$,  we employ a modified recursion for the displacements,
$x_t=|x_{t-1}+\xi_t|$ while the (externally given) fractional Gaussian noise remains unchanged \cite{JeonMetzler10}.
This means if the particle's position $x_t$
happens to be negative, it is placed at $-x_t$ instead.
Alternatively, one could, e.g.,
place the particle at the origin via $x_t=\max(x_{t-1}+\xi_t,0)$.
Both versions should yield the same long-time behavior because individual
steps have a finite characteristic length of $\sigma$. Indeed, we have numerically confirmed that
their results agree for times fulfilling $\langle x_t^2 \rangle \gg \sigma^2$.

To set the stage, let us briefly summarize reflected \emph{normal} Brownian
motion ($\alpha=1$). The probability density $P(x,t)$ of the particle position can be
found by solving the diffusion equation $\partial_t P = (\sigma^2/2)\partial_x^2 P$
under the flux-free boundary condition $\partial_x P=0$ at $x=0$ and initial condition
$P(x,0) =\delta(x)$. This yields the Gaussian
\begin{equation}
P(x,t)= \sqrt{\frac 2 {\pi \sigma^2 t}} \exp \left( -\frac {x^2}{2\sigma^2 t} \right),
\label{eq:half-Gaussian}
\end{equation}
restricted to nonnegative $x$-values. The mean-square displacement
$\langle x_t^2 \rangle$ thus increases as $t$, just as in the unconfined case.
Importantly, for normal Brownian motion, the reflecting
wall does not change the Gaussian character of $P(x,t)$.

\paragraph*{Monte Carlo simulations.}

We perform simulations of the discrete-time reflected FBM for
anomalous diffusion exponents $\alpha$ ranging from 0.4 to 1.8. Each simulation
uses up to $5\times 10^7$ particles that start from the origin and perform up to
$6.7\times 10^7$ ($2^{26}$) time steps. The correlated Gaussian
random numbers representing the fractional noise $\xi_i$ are generated by means of the
Fourier-filtering method \cite{MHSS96}. It starts from a sequence of independent
Gaussian random numbers $\chi_i$. The Fourier transform $\tilde \chi_\omega$ of these numbers
is then converted via ${\tilde{\xi}_\omega} = [\tilde C(\omega)]^{1/2} \tilde{\chi}_\omega$,
where $\tilde C(\omega)$ is the Fourier transform of the correlation function (\ref{eq:xi_corr}).
The inverse Fourier transformation of the ${\tilde{\xi}_\omega}$ gives the desired noise values.
In our simulations, the variance $\sigma^2$ of the $\xi_i$ is fixed at unity.

\begin{figure}[t]
\includegraphics[width=\columnwidth]{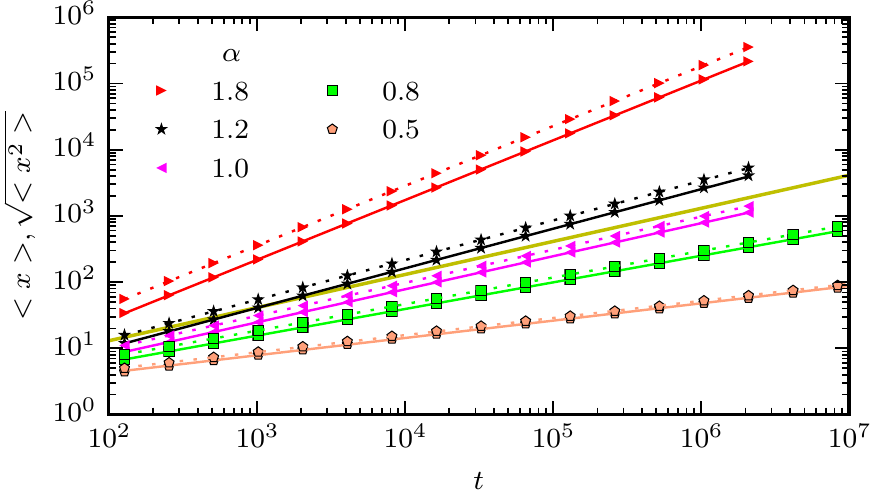}
\caption{Average (solid lines) and root-mean-square (dashed lines) displacements of a reflected random
  walker vs.\ time $t$ for several values of the anomalous diffusion exponent $\alpha$. The relative statistical errors
  of the data are about $10^{-2}$, much smaller than the symbol size. The solid and dashed lines represent power-law fits.
  The thick yellow line shows normal diffusion behavior, $\langle x^2 \rangle \sim t$, with arbitrary prefactor.}
\label{fig:meansquare}
\end{figure}

Figure \ref{fig:meansquare} shows the resulting time dependencies of the average displacement $\langle x_t \rangle$
and the root-mean-square displacement $\langle x_t^2 \rangle^{1/2}$
for several values of the anomalous  diffusion
exponent $\alpha$ used to create the fractional noise $\xi_i$.
The figure demonstrates that the mean-square displacement $\langle x_t^2 \rangle$
increases as $t^\alpha$, just as in the unconfined case. Power-law fits yield exponent values of
1.806(10), 1.196(6), 0.998(6), 0.804(4), and 0.51(2) for $\alpha=1.8$, 1.2, 1.0, 0.8, and 0.5, respectively.
(The numbers in brackets give the error of the last digit.)
As the barrier restricts the motion to nonnegative $x$-values, the average
displacement $\langle x_t \rangle$ is nonzero and increases as $t^{\alpha/2}$.

\begin{figure}
\includegraphics[width=\columnwidth]{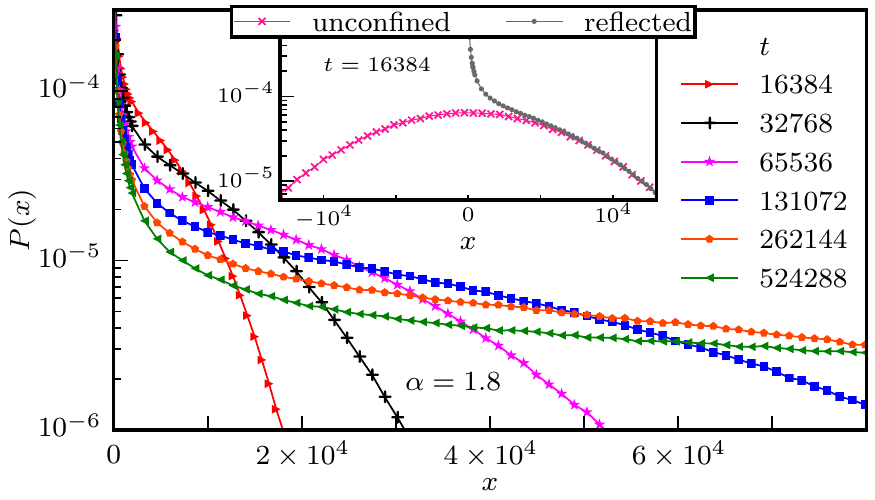}
\caption{Probability density function $P$ of the position $x$ at different times $t$ for $\alpha=1.8$.
  The statistical errors of the data are
  smaller than the symbol size. A comparison of $P(x)$ for unconfined and reflected FBM is shown in the inset.}
\label{fig:SubSuper_diffusive}
\end{figure}

While the average and mean-square displacements of the particle show
the expected behavior, the probability density $P(x)$ of its position
displays surprising features. The probability density of unconfined FBM is a Gaussian.
Based on the results for reflected normal Brownian motion,
one might expect that $P(x)$ for reflected FBM is a Gaussian of the appropriate width and
restricted to nonnegative $x$ values. However, Fig.\ \ref{fig:SubSuper_diffusive}
demonstrates striking deviations from Gaussian behavior for the example of $\alpha=1.8$.
Specifically, particles accumulate close to the reflecting wall.
This creates a divergence of $P(x)$ for $x \to 0$ while the large-$x$ behavior remains
Gaussian. We observe analogous behavior for all
$\alpha$ in the superdiffusive regime ($\alpha=1.1$ to 1.8).
For subdiffusive $\alpha$ (0.4 to 0.9), in contrast, the region close to
the reflecting wall is depleted of particles, and $P(0)$ approaches zero.
For $\alpha=1$, our data agree with the half-Gaussian (\ref{eq:half-Gaussian}) resulting from
the solution of the normal diffusion equation.

\begin{figure}
\includegraphics[width=\columnwidth]{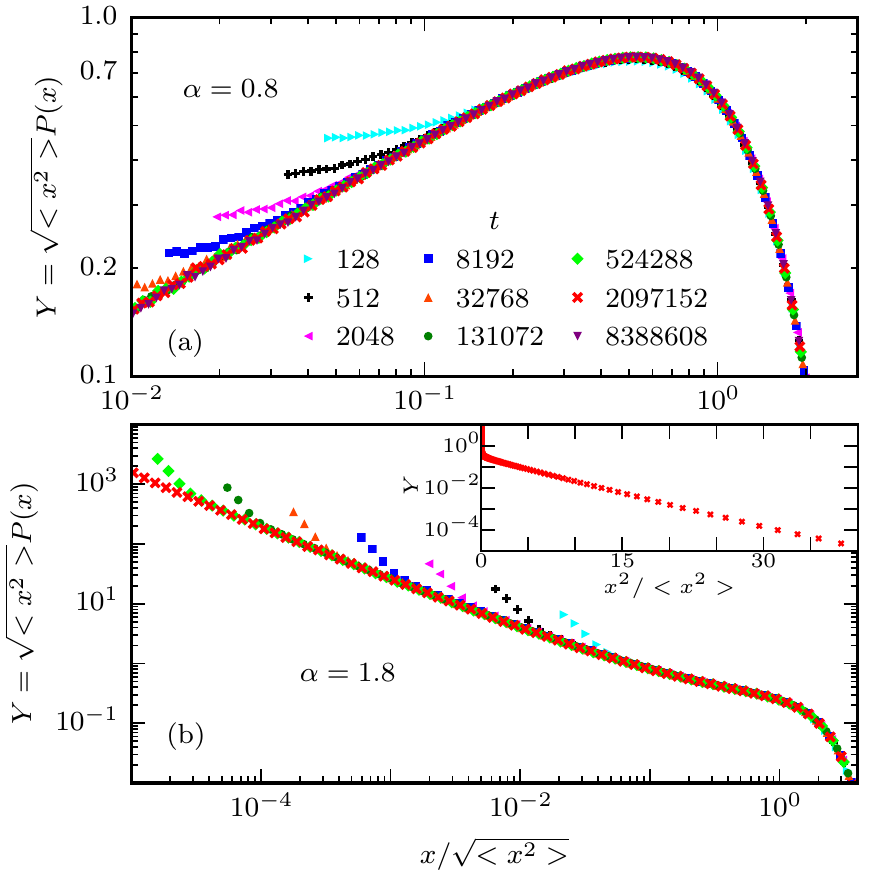}
\caption{Scaled probability density  $Y=\langle x^2\rangle^{1/2} P$ vs.\ $x/\langle x^2\rangle^{1/2}$
  for $\alpha=0.8$ [panel (a)] and 1.8 [panel (b)] for several times $t$.
  The probability densities for different $t$ collapse onto a common master curve. The deviations at small $x$
  for early times stem from the discrete-time character of our simulations (see text).
  The inset shows a log-linear plot of $Y$ vs.\ $x^2$, demonstrating the Gaussian character of the large-$x$ tail.}
\label{fig:ScaledDistr}
\end{figure}

Despite the non-Gaussian character, the probability densities at different times can be scaled
to fall onto a common master curve if they are expressed in terms of $y=x/\langle x_t^2 \rangle^{1/2}=x/(\sigma t^{\alpha/2})$.
This is illustrated in Fig.\ \ref{fig:ScaledDistr} for the examples of $\alpha=1.8$ (superdiffusive)
and $\alpha=0.8$ (subdiffusive).
The scaling collapse means that the probability density can be written in the form
\begin{equation}
P(x,t) = \frac 1 {\sigma t^{\alpha/2}}\, Y\left( x/(\sigma t^{\alpha/2}) \right )
\label{eq:P_scaling_t}
\end{equation}
where $Y$ is a dimensionless scaling function. We  observe analogous scaling behavior for
all investigated $\alpha$. (For normal Brownian motion, $\alpha=1$, it follows directly from eq.\
(\ref{eq:half-Gaussian}).)
This implies that the singularity observed close to $x=0$ is not a finite-time artefact but
part of the (asymptotic) long-time behavior.
Note that the deviations from the scaling form
appearing in Fig.\ \ref{fig:ScaledDistr} for small displacements $x$ at short times $t$
arise because we use a discrete-time version of FBM. The scaling form only holds for
$x \gg \sigma$ as the Gaussian distributed step $\xi$ obscures the structure
of $P(x,t)$ for $x \lesssim \sigma$.

\begin{figure}
\includegraphics[width=\columnwidth]{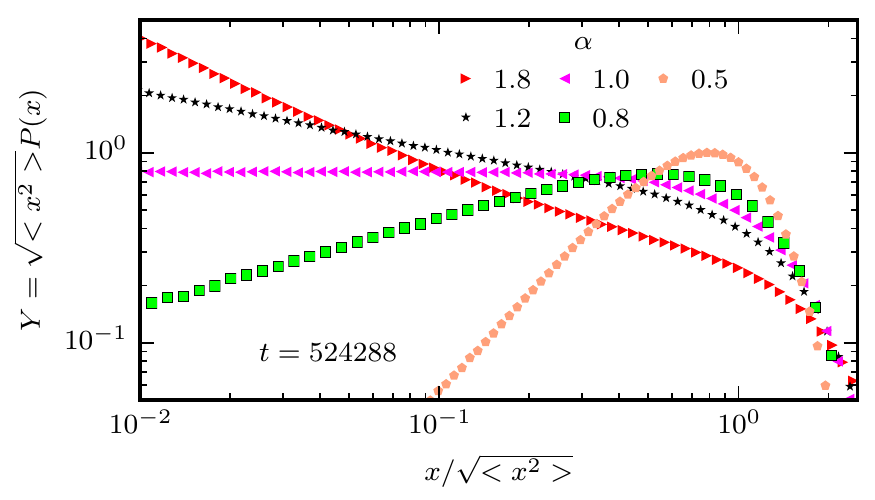}
\caption{Scaled probability density function, $\langle x^2\rangle^{1/2} P$ vs.\ $x/\langle x^2\rangle^{1/2}$,
  at $t=524288$ for several values of $\alpha$. For small $x$, the probability densities follow power laws
  $P(x) \sim x^\kappa$. Power law fits of the small-$x$ behavior yield
  $\kappa \approx -0.89$, -0.33, 0.00, 0.47, and 1.75 for $\alpha=1.8$, 1.2,
  1.0, 0.8, and 0.5, respectively.}
\label{fig:ScaledDistr_gamma}
\end{figure}

Let us now analyze  in more detail the functional form of the singularity of the probability density function
$P(x)$ for $x \to 0$. Figure \ref{fig:ScaledDistr_gamma} presents a double-logarithmic
plot of the scaled probability densities at time $t=524288$ for several values of $\alpha$.
All curves become straight lines at small $x$, i.e., they feature power-law behavior $P(x) \sim x^\kappa$.
(For $\alpha=1.8$, the power law
at small $x$ is preceded by a wide crossover region.
The simulations thus require long times to access the asymptotic small-$x$ regime.)

\begin{figure}
\includegraphics[width=\columnwidth]{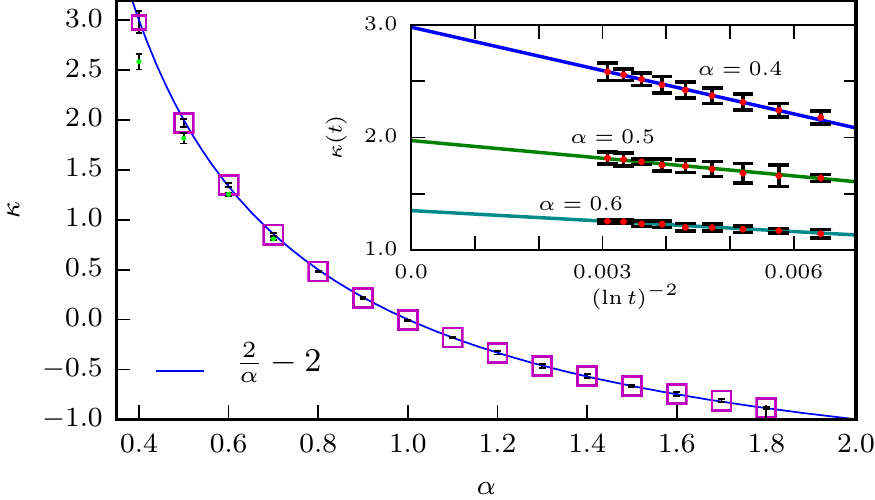}
\caption{Exponent $\kappa$ vs.\ anomalous diffusion exponent $\alpha$.
 The $\kappa$ values stem from power-law fits of the small-$x$ behavior of $P(x,t)$
 at the longest times ($t=2\times 10^6$ to $6.7\times 10^7$ depending on $\alpha$).
 The error bars combine the statistical error and the uncertainty of
 the fit interval. The solid line represents the conjectured function $\kappa=2/\alpha -2$.
 For $\alpha \le 0.7$, the squares mark the extrapolated (to infinite time) $\kappa$ values
 while the (green) dots show the effective $\kappa$ at the longest simulation time.}
\label{fig:Kappa_alpha}
\end{figure}

To determine the singularity exponent $\kappa$ accurately, we employ power-law fits of the small-$x$
behavior of $P(x,t)$ obtained at the longest times $t$.
Figure \ref{fig:Kappa_alpha} shows the resulting dependence of $\kappa$ on the
anomalous diffusion exponent $\alpha$.
The data indicate that $\kappa$ decreases monotonically with $\alpha$.  In the subdiffusive regime,
$\alpha<1$, it takes positive values (such that $P(x)$ vanishes at $x=0$). For normal Brownian
motion, $\alpha=1$, we find $\kappa=0$ which implies that $P(x)$ approaches a constant for $x\to 0$.
This agrees with the analytical solution (\ref{eq:half-Gaussian}). In the superdiffusive case, $\alpha>1$,
the exponent $\kappa$ is negative, corresponding to a divergence of $P(x)$ at $x=0$.
Note that the $\kappa$ values obtained from the fit show a significant dependence
on the simulation time for $\alpha \le 0.7$, indicating a slow crossover to the asymptotic behavior. We therefore extrapolate
these values to infinite time, as shown in the inset of Fig.\ \ref{fig:Kappa_alpha}
\footnote{The $(\ln t)^{-2}$ dependence in the inset of Fig.\ \ref{fig:Kappa_alpha}
was chosen empirically such that the data lie on a straight line.}.

The exact functional form of the $\kappa(\alpha)$ dependence is not known. We find, however,
that the empirical function $\kappa=2/\alpha-2$ describes the data well. In fact, the agreement is
excellent over the entire $\alpha$ range, if we use the extrapolated $\kappa$ values for $\alpha \le 0.7$.
In the limit $\alpha \to 0$, the function $\kappa=2/\alpha-2$ predicts $\kappa$ to diverge.
In the ballistic limit, $\alpha \to 2$, the function  predicts $\kappa \to -1$.
As a power law with $\kappa=-1$ is not normalizable, this means the singularity turns into
a $\delta$-peak.

The probability density for ballistic motion ($\alpha=2$, where the $\xi_i$ are perfectly correlated in time)
can actually be found analytically. Half of the particles (those with negative $\xi_1$)
get stuck at the wall forever while those with positive $\xi_1$ move to the right with
Gaussian-distributed speeds. $P(x)$ is thus a sum of a half-Gaussian of width $\sigma t$
and a delta-peak (of weight 0.5) at the origin, in agreement with the ballistic limit of the
conjectured $\kappa(\alpha)$ function.

\paragraph*{Discussion.}

\begin{figure}
\includegraphics[width=\columnwidth]{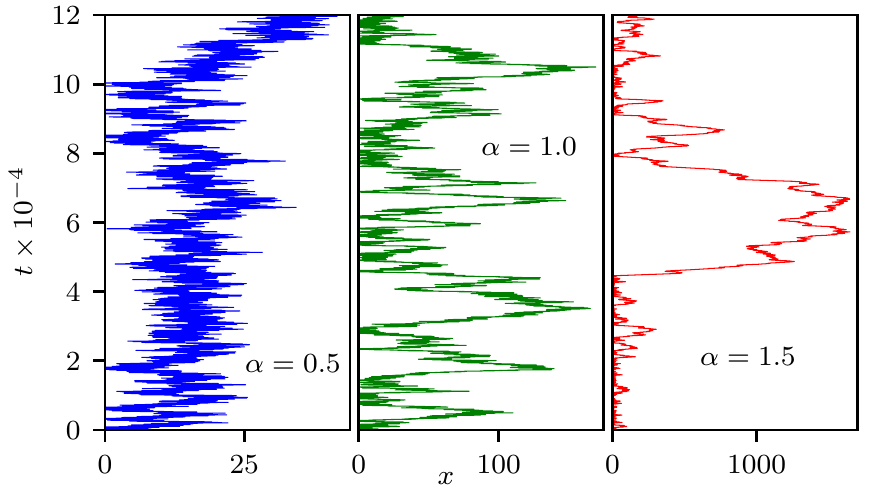}
\caption{Sample trajectories of reflected FBM for the cases of subdiffusion
($\alpha=0.5$), normal diffusion ($\alpha=1.0$), and superdiffusion ($\alpha=1.5$).
The long-range correlations of the FBM steps cause the particle to get stuck at the wall (at $x=0$)
for long times in the superdiffusive case. In contrast, the particle will tend to move away from
the wall right after reaching it for subdiffusion.}
\label{fig:Trajectories}
\end{figure}

In summary, our central result is the striking non-Gaussian behavior of reflected FBM,
caused by the interplay between the boundary condition and the long-range correlations.
The probability density $P(x)$ exhibits a power-law singularity, $P(x) \sim x^\kappa$, at the barrier.
It can be understood qualitatively as follows.
For persistent correlations (superdiffusion), the particle will attempt to continue in the negative $x$-direction
upon reaching the wall. As the wall prevents this, the particle will get stuck at the wall for
a long time \footnote{The probability of finding long periods of motion in predominantly one direction
is discussed in Ref.\ \cite{IbrahimBarghathiVojta14} for power-law correlated disorder.},
increasing the probability density there. For antipersistent correlations (subdiffusion),
in contrast, the particle will tend to move away from the wall right after reaching it, reducing the probability
density compared to the uncorrelated (normal diffusion) case. This is illustrated in Fig.\
\ref{fig:Trajectories}.

We note in passing that non-Gaussian fluctuations of individual trajectories
were recently discovered in superdiffusive FBM \cite{SchwarzlGodecMetzler17}.
Moreover, non-Gaussian behavior in diffusive dynamics can also be caused by several other mechanisms \cite{MetzlerKlafter00,HoeflingFranosch13,MJCB14,WangKuoBaeGranick12,ChubynskySlater14,Jeonetal16,MatseChubynskyBechhoefer17,ChechkinSenoMetzlerSokolov17,Lampoetal17}.
More generally, long-range correlations and the corresponding nonanalyticities
can arise even for normal diffusion in the presence of soft modes or quenched
disorder \cite{AlderWainwright67,LeeuwenWeijland67,FranoschHoflingBauerFrey10}.

It is instructive to compare our results for FBM with the behavior
of another anomalous diffusion model called scaled Brownian motion (SBM) \cite{LimMuniady02,JeonChechkinMetzler14}.
SBM can be understood as
normal diffusion with a time-dependent diffusion constant.
Its probability density  fulfills the generalized diffusion equation
$\partial_t P = \alpha  t^{\alpha-1}   (\sigma^2/2) \partial_x^2 P$. In the unconfined case
(no barrier), the resulting probability density of a particle starting
at the origin is a Gaussian of zero mean and variance $\langle x_t^2\rangle = \sigma^2 t^\alpha$.
This means, it is identical to the probability density of FBM. However, in contrast to FBM,
the probability density of SBM remains Gaussian in the presence of a reflecting wall. This follows
from the fact that the Gaussian fulfills not only the generalized diffusion equation
but also the flux-free boundary condition $\partial_x P=0$ imposed by the barrier.

Reflected random walks find numerous applications in physics, chemistry, biology and beyond.
The singularity at $x=0$ in the probability density function $P(x)$ of the particle position
will be particularly important in applications that are dominated by rare events. Imagine, for example,
that one is interested in a quantity $z=e^{-x}$ that depends exponentially on the
position (see Ref.\ \cite{VojtaHoyos15} for a recent example
of such a situation). The average of $z$ is dominated by particles close to the origin.
Indeed, a straightforward calculation shows that $\langle z \rangle \sim t^{-\alpha(1+\kappa)/2}$
for sufficiently long times. The appearance of $\kappa$ in this relation
means that the singularity in $P(x)$ affects the
long-time behavior qualitatively.

How robust are our results if the correlation function $C(j)$ of the steps is modified?
If the correlations are persistent (positive) and
$C(j)$ decays for large $j$ as a power law $j^{\alpha-2}$ with $\alpha >1$ (i.e., more slowly than $j^{-1}$),
the resulting long-time behavior is expected to be identical to the corresponding FBM. This implies
superdiffusive motion and a divergence of $P(x)$ at the origin. We have confirmed this
by simulations using $C(j)=(1+j^2)^{(\alpha-2)/2}$. In contrast, for persistent
correlations that decay faster than $j^{-1}$, the behavior is expected to agree
with that of normal uncorrelated Brownian motion.
The subdiffusive behavior occurring for FBM with $\alpha<1$ is more fragile as it
relies on the antipersistent correlations fulfilling $\sum_j C(j)=0$.
A generic antipersistent correlation function that instead fulfills
$\sum_j C(j)=\textrm{const}\ne 0$ is expected to produce normal diffusion behavior.

So far, we have considered unbiased reflected FBM. It also interesting to ask,
how the reflecting wall influences the biased case. If the bias is away from the barrier
(in the positive $x$-direction),  the barrier will become less important with
increasing time. For long times we thus expect to recover the Gaussian probability density
of unconfined FBM. For bias towards the barrier (in negative $x$-direction), in contrast,
we expect a steady state whose probability density is determined by the interplay of
the long-range correlations and the bias.

To conclude, the interplay between the geometric confinement and the long-time memory
encoded in the FBM correlations leads to highly non-Gaussian behavior with a singular
probability density. The mechanism causing the singularity appears to
be general; we thus expect our results to provide a framework for a large class of
long-range correlated processes in nontrivial geometries.

\begin{acknowledgments}
This work was supported by the NSF under Grant Nos. PHY-1125915 and DMR-
1506152 and by the S\~ao Paulo Research Foundation (FAPESP) under Grant No. 2017/08631-0.
We thank Ralf Metzler for valuable discussions.
T.V. is grateful for the hospitality of the Kavli Institute
for Theoretical Physics, Santa Barbara where part of the research was performed
\end{acknowledgments}

\bibliographystyle{apsrev4-1}
\bibliography{../00Bibtex/rareregions}
\end{document}